\documentclass{PoS}
\usepackage{amsmath,slashed,subfigure,setspace}
\pdfoutput=1

\newcommand{\mps}{\ensuremath{m_{\text{PS}}}}
\newcommand{\mv}{\ensuremath{m_{\text{V}}}}
\newcommand{\ms}{\ensuremath{m_{\text{S}}}}

\newcommand{\mpv}{\ensuremath{m_{\text{PV}}}}
\newcommand{\msh}{\ensuremath{m_{1/2}}}
\newcommand{\mgb}{\ensuremath{m_{0^{++}}}}

\newcommand{\shalf}{\ensuremath{\text{spin 1/2}}}
\newcommand{\su}[1]{\ensuremath{\text{SU}(#1)}}

\newcommand{\bpsi}{\bar{\psi}}
\DeclareMathOperator{\Tr}{Tr}
\DeclareMathOperator{\tr}{tr}

\title{Spectrum and mass anomalous dimension of\\ SU(2) gauge theories
   with fermions in the adjoint\\ representation: from $\bf {N_f=1/2}$ to
   $\bf {N_f=2}$}

\ShortTitle{Lattice simulations of adjoint QCD}

\author{\speaker{Georg Bergner}\\
Friedrich-Schiller-University Jena, Institute of Theoretical Physics,\\
Max-Wien-Platz 1, D-07743 Jena, Germany\\
E-mail: \email{georg.bergner@uni-jena.de}}

\author{Pietro Giudice, Gernot M\"unster\\
Universit\"at M\"unster, Institut f\"ur Theoretische Physik,\\
Wilhelm-Klemm-Str. 9, D-48149 M\"unster, Germany\\
E-mail: \email{p.giudice@uni-muenster.de, munsteg@uni-muenster.de}}

\author{Istvan Montvay\\
Deutsches Elektronen-Synchrotron DESY,\\
Notkestr. 85, D-22603 Hamburg, Germany\\
E-mail: \email{montvay@mail.desy.de}}

\author{Stefano Piemonte\\
Universit\"at Regensburg, Institute for Theoretical Physics,\\
D-93040 Regensburg, Germany\\
E-mail: \email{stefano.piemonte@ur.de}}

\abstract{%
We summarize our results concerning the spectrum and mass anomalous dimension of SU(2) gauge theories with various numbers of fermions in the adjoint representation, 
where each Majorana fermion corresponds effectively to half a Dirac flavour $N_f$.  The most relevant examples for extensions of the standard model are supersymmetric Yang-Mills theory ($N_f=1/2$)
and Minimal Walking Technicolour ($N_f=2$).  In addition to these theories we will also consider the cases of $N_f=1$ and $N_f=3/2$. 
The results comprise the particle spectrum of glueballs, triplet and singlet mesons, and possible fractionally charged spin half particles. In addition we will discuss our recent results for the mass anomalous dimension.}

\FullConference{The 34rd International Symposium on Lattice Field Theory\\
                           24-30 July 2016\\
                          University of Southampton, UK}

\begin{document}
\section{Technicolour and adjoint QCD}

Gauge theories with fermions in higher representations of the gauge group offer  interesting alternatives for extensions of the standard model
of particle physics and for a more general view on theoretical problems like the understanding of the confinement mechanism. One of the most interesting
examples in this context is the adjoint representation, since it is related to several approaches to extensions of the standard model, in particular supersymmetry 
and Technicolour models. They represent theories that are in some respects quite different from QCD-like examples with fermions in the fundamental representation.
From that perspective they are interesting from the more general theoretical point of view.

In this work we present our investigations related to the Walking Technicolour approach to extensions of the standard model, see \cite{Hill:2002ap,
Piai:2010ma} for a review. In this approach a more
natural electroweak sector is obtained from a new strongly coupled dynamics. The Higgs particle appears as a bound state in this strongly coupled theory. There are severe constraints on 
Technicolour candidates, since they have to be consistent with the electroweak precision data and still potentially explain the fermion masses based on an extended Technicolour theory. Furthermore
a large mass hierarchy between the Higgs particle and so far undiscovered additional bound states has to be explained. One possible explanation is a dynamics that is quite different from QCD
due to a near conformal walking behaviour. This means that such kind of theories have to be close to the conformal window. 

The conformal window is defined by the appearance of an infrared fixed point of the running gauge coupling. In the perturbative calculations the running of the gauge coupling of the confining 
gauge theory gets weaker with an increasing number of fermions $N_f$. Above a certain $(N_f)_c$ the running stops at an infrared fixed point. Increasing the number of fermions even further, asymptotic 
freedom is lost. The conformal window is defined as the region between $(N_f)_c$ and the loss of asymptotic freedom. 

The determination of the conformal window is of general theoretical interest since it marks the parameter regions for a completely different behaviour of the theory in the infrared. Irrespective of 
the considered application, it is important to know this behaviour. While the upper end of the conformal window is in the perturbative regime, the lower end can only be determined by non-perturbative methods.

There have been several investigations of the conformal window for fermions in the fundamental representation in a scan of several theories with different $N_f$, see \cite{Appelquist:2016viq,DeGrand:2015zxa,Nogradi:2016qek} for a review of lattice results.
In our investigations we consider the same 
approach for a determination of the conformal window in the case of fermions in the adjoint representation. Higher fermion representations allow to approach the lower end of the conformal 
window with a smaller number of fermions $(N_f)_c$. This allows to construct theories that might have less tensions with the electroweak precision data.

In our numerical investigations on the lattice, a finite mass term is included in the theory, which leads to mass deformed versions of the conformal candidates. In this case the mass is the only 
relevant parameter at the fixed point, which implies that for small $m$ all particle masses scale like $M\sim m^{1/(1+\gamma_m)}$ with the mass anomalous dimension $\gamma_m$ at the fixed point. 
It also implies that a signal for conformality are the constant ratios of different particle masses. The value of $\gamma_m$ can be obtained from the scaling of the mass ratios. A more precise 
measurement is provided by the mode number, the integrated eigenvalue density of the Dirac operator \cite{Giusti:2002sm,Patella:2012da}. From the perspective of the Walking Technicolour approach, a large anomalous dimension is desirable. 
It is therefore interesting to study the $N_f$ dependence of this quantity.

$N_f=2$ \su{2} adjoint QCD, also called Minimal Walking Technicolour (MWT), has been studied in several investigations \cite{Catterall:2007yx,Hietanen:2009zz,DelDebbio:2010hx,DeGrand:2011qd,Rantaharju:2015yva,DelDebbio:2015byq}. 
It shows the signatures of a conformal theory and a clear separation between 
the lightest scalar bound state and the rest of the spectrum. The case of $N_f=1/2$, corresponding to \su{2} $\mathcal{N}=1$ supersymmetric Yang-Mills theory, has been investigated by us in several studies. 
This theory clearly shows the signals of a theory below the conformal window. In more recent investigations the $N_f=1$ case has been considered. In this short report we add our results for $N_f=2$ 
and $N_f=3/2$ and provide a summary of the properties of the different theories with fermions in the adjoint representation.

In our studies of supersymmetric Yang-Mills theory the investigation of different inverse gauge couplings $\beta$ turned out to be important. In a QCD-like theory this corresponds to a study 
of the approach towards the continuum limit. In the conformal case this parameter is irrelevant in the near vicinity of the fixed point. Since we want to investigate all the theories on the same ground, we include 
a study of the $\beta$-dependence in our investigations. 

\section{The expected mass spectrum for adjoint QCD}

Adjoint QCD in the continuum is defined by the following action
\begin{align}
 \mathcal{L}=
 \Tr\left[-\frac{1}{4}F_{\mu\nu}F^{\mu\nu}
 +\sum_{i=1}^{N_f}\bpsi_{i}(\slashed{D}+m)\psi_{i}\right],
\end{align}
with the covariant derivative for fermions in the adjoint representation
\begin{align}
D_\mu \psi =\partial_{\mu}\psi + \mathrm{i} g [A_{\mu},\psi]\; .
\end{align}
$N_f$ counts the number of Dirac fermions, which corresponds to $2N_f$ Majorana fermions, since the adjoint representation is 
compatible with the Majorana condition $\psi=C\bpsi^T$.

The chiral symmetry breaking pattern, which is different from QCD, is implied by the representation in terms of Majorana flavours.
In adjoint QCD there is a larger chiral symmetry group with the spontaneous breaking pattern
\begin{align}
 \su{2N_f} \rightarrow \text{SO}(2N_f)
\end{align}
in the presence of a fermion condensate.
The chiral symmetry breaking leads to the appearance of light pseudo-Nambu-Goldstone particle states. A signal for these states can be obtained by the pseudoscalar triplet meson,
similar to the pion in QCD. In our current work this adjoint pion mass is represented by the pseudoscalar mass $\mps$.
In our simulations we employ a discretisation of the theory with a Wilson-Dirac operator including stout-smeared link variables and a tree-level Symanzik-improved gauge action.

We study various particle states in the theory. One class are the flavour triplet mesons: the pseudoscalar \mps, scalar \ms, vector \mv, and pseudovector  \mpv\ mesons. 
We have also considered the flavour singlet scalar and pseudoscalar mesons. In addition we study the scalar $0^{++}$ glueball, providing a signal for a possible light Higgs-like state. 
As an example for more exotic fermion-gluon operators that are possible in adjoint QCD, we measure the mass obtained from the correlator of the  spin-1/2 operator
\begin{align}
  \sum_{\mu,\nu} \sigma_{\mu\nu} \tr\left[F^{\mu\nu} \psi \right] .
\end{align}
\section{Results for Minimal Walking Technicolour including the $\beta$ dependence}
\begin{figure}
\centering
\subfigure[Masses in units of $\mps$ at $\beta=1.5$]{ \includegraphics[width=0.49\textwidth]{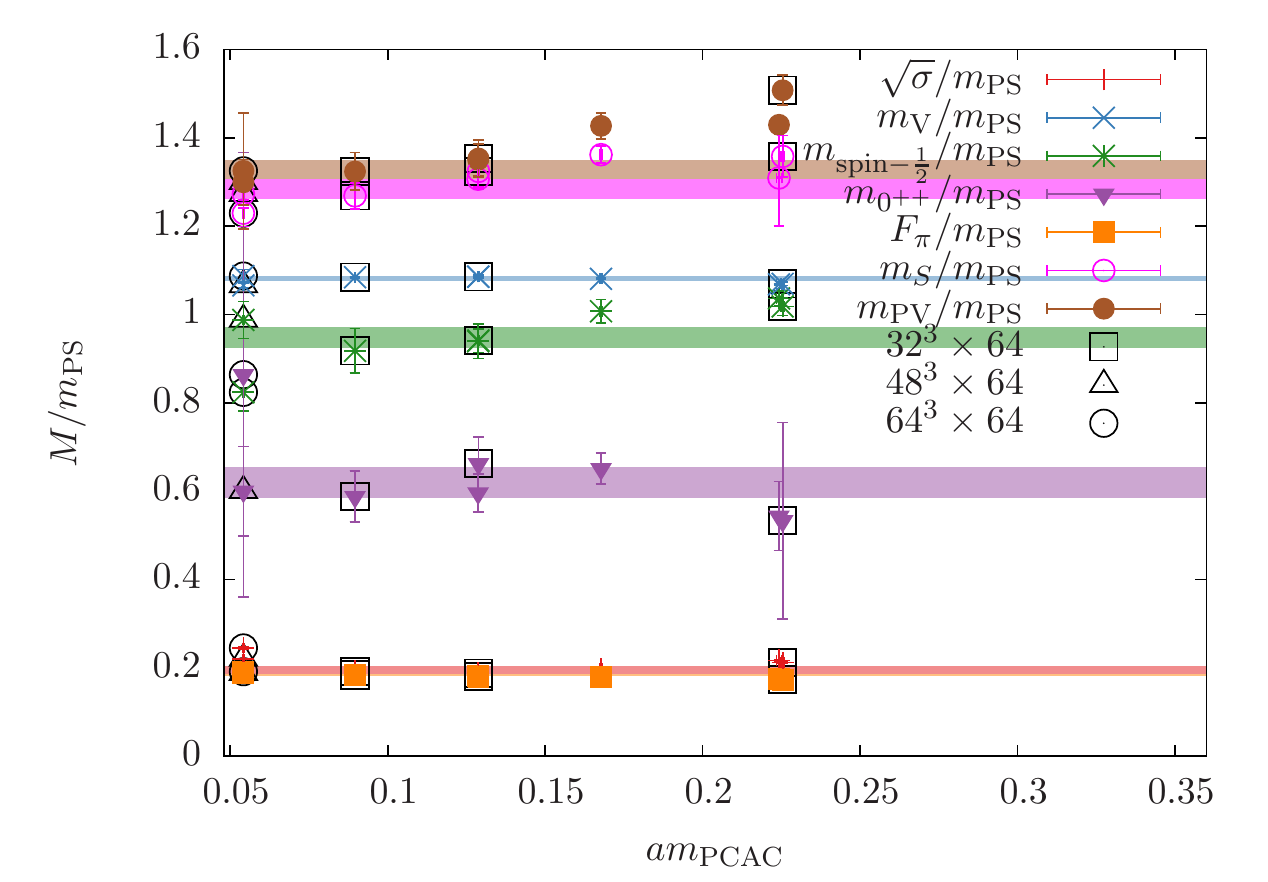}\label{ratiosb150}}
\subfigure[Masses in units of $\mps$ at $\beta=1.7$]{ \includegraphics[width=0.49\textwidth]{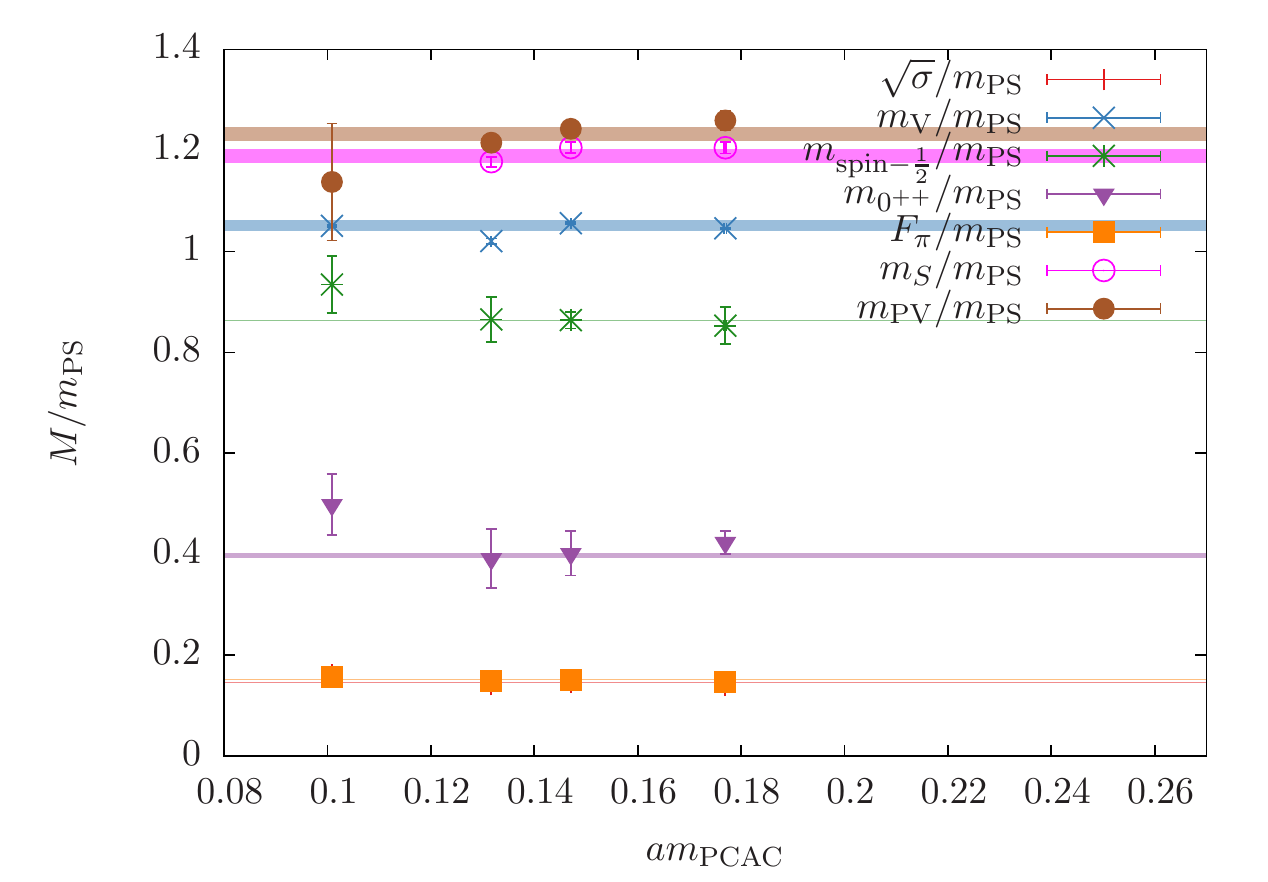}\label{ratiosb170}}
 \caption{The masses of various bound states in MWT in units of the pseudoscalar mass $\mps$. 
 The triplet mesons in the scalar ($\ms$), vector ($m_V$), and pseudovector ($m_{PV}$) channel are shown together with the scalar glueball ($m_{0^{++}}$) and the $\shalf$\ particle ($m_{\shalf}$).
 The figure also shows the pseudoscalar decay constant $F_\pi$ and the string tension $\sigma$.\label{mwtmass}
 }
\end{figure}
\begin{table}
 \begin{center}
  \begin{tabular}{ |c|c | c|}
 \hline 
 State & $\beta=1.5$ & $\beta=1.7$ \\
  \hline
  \mv  &1.0825(58) & 1.051(12) \\
  \ms  & 1.285(24)  & 1.190(14)  \\
  \mgb &  0.620(35) &  0.398(48)\\
  \msh          &  0.948(24)   & 0.86394(52)\\
  \hline
 \end{tabular}
 \end{center}
\caption{The mass ratios averaged over a range of PCAC masses for two different $\beta$ values. The values correspond to the masses of the triplet vector (\mv) and triplet scalar (\ms) mesons,
the glueball (\mgb), and the spin-1/2 particle (\msh) divided by the triplet pseudoscalar meson mass (\mps). }
\label{tab:ratio}
\end{table}
\begin{table}
 \begin{center}
 \begin{tabular}{ |c|c | c| c|  }
 \hline 
$N_s\times N_t$ & $\beta$ & $\kappa$ & $\gamma_\ast$\\
  \hline 
  $24\times 64$ &1.5 & 0.1325  & 0.39(3) \\
  $32\times 64$ &1.5 & 0.1335  & 0.38(1) \\
  $48\times 64$ &1.5 & 0.1344  & 0.380(10) \\
  $32\times 64$ &1.5 & 0.1350  & 0.375(4) \\
  average           &1.5&&0.376(3)\\
  \hline 
  $32\times 64$ &1.7 & 0.1285  & 0.270(15) \\
  $32\times 64$ &1.7 & 0.1290  & 0.260(20) \\
  $32\times 64$ &1.7 & 0.1300  & 0.285(15)\\
  average            &1.7&&0.274(10)\\
  \hline
 \end{tabular}
 \end{center} 
\caption{This Table contains the estimates for the mass anomalous dimension $\gamma_\ast$ at the fixed point, that we have obtained from a fit of the mode number.
The results are obtained at two different values of $\beta$.}
\label{tab:gamma}
\end{table}
Our investigation of Minimal Walking Technicolour has now been finalised. We have measured the spectrum and determined the mass anomalous dimension from the mode number. 
The general picture of our results is consistent with earlier investigations \cite{DelDebbio:2010hx,DelDebbio:2015byq}: the particle spectrum contains a light scalar particle, namely the $0^{++}$ glueball state, while
the triplet mesons, including the pseudoscalar meson, are heavier. Hence the spectrum in not consistent with chiral symmetry breaking, which predicts the appearance of light pseudo-Goldstone 
particles separated from the rest of the spectrum. The mass ratios are approximately constant as a function of the PCAC mass, see Figure \ref{ratiosb150} and 
\ref{ratiosb170}.

In our more detailed investigations we have also added some new states that have not been considered in earlier investiations: the fermionic spin-1/2 particle, and the flavour singlet mesons.
The fermion-gluon particle is lighter than the pseudoscalar meson, but considerable heavier than the scalar glueball. This observation is relevant from a phenomenological piont of few, since 
these states might lead to fractionally charged particles.

We obtained results at two different $\beta$-values that are superficially consistent with each other, as shown in Figure \ref{ratiosb150} and 
\ref{ratiosb170}. The same ordering of the states with constant mass ratios is observed. A more careful investigation of the mass ratios shows, 
however, a significant $\beta$ dependence of the results. We have averaged the mass ratios over a certain range of the PCAC mass for both values of $\beta$.
As shown in Table \ref{tab:ratio}, the results for both $\beta$ values are not consistent with each other. In particular the scalar glueball at the larger 
value of $\beta$ is lighter in comparison to $\mps$ than at $\beta=1.5$. This indicates a remnant $\beta$-dependence and 
scaling corrections for the conformal theory.

We have estimated the mass anomalous dimension $\gamma_\ast$ at the fixed point from our measurement of the mode number. The mode number corresponds to the integrated
spectral density of the Wilson-Dirac operator. Our results at $\beta=1.5$ are consistent with \cite{DelDebbio:2015byq} ($\gamma_\ast=0.371(20)$), but for  
$\beta=1.7$ a considerable smaller $\gamma_\ast$ is obtained, see Table \ref{tab:gamma}. The remnant scaling corrections that are indicated by these results might also explain the discrepancies between 
different results in the literature, for example the small $\gamma_\ast=0.20(3)$ obtained with a clover improved Dirac operator in \cite{Rantaharju:2015yva}.
\section{New results for $N_f=3/2$}
\begin{figure}
\centering
\subfigure[Masses in lattice units]{\includegraphics[width=0.48\textwidth]{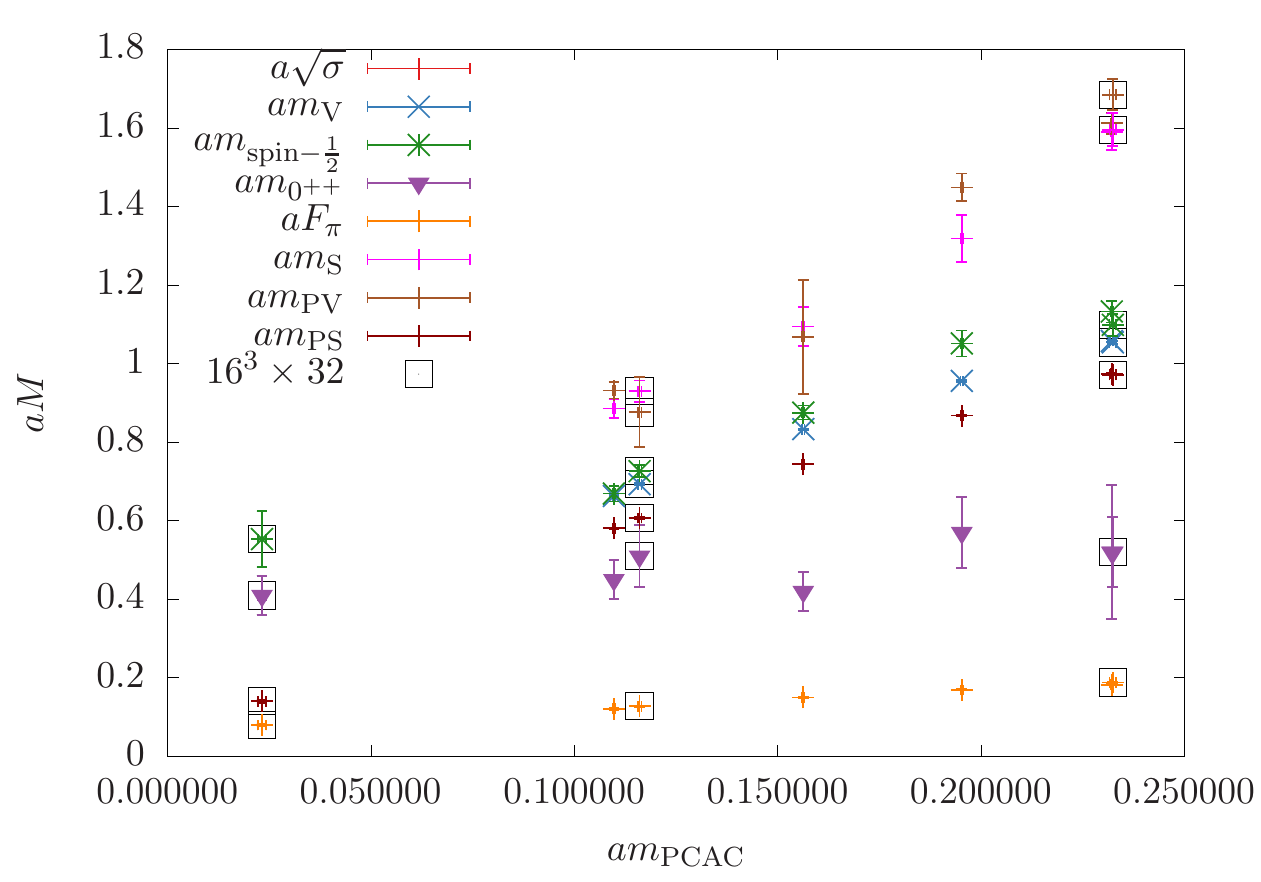}\label{massesnf32}}
\subfigure[Masses in units of $\mps$]{ \includegraphics[width=0.48\textwidth]{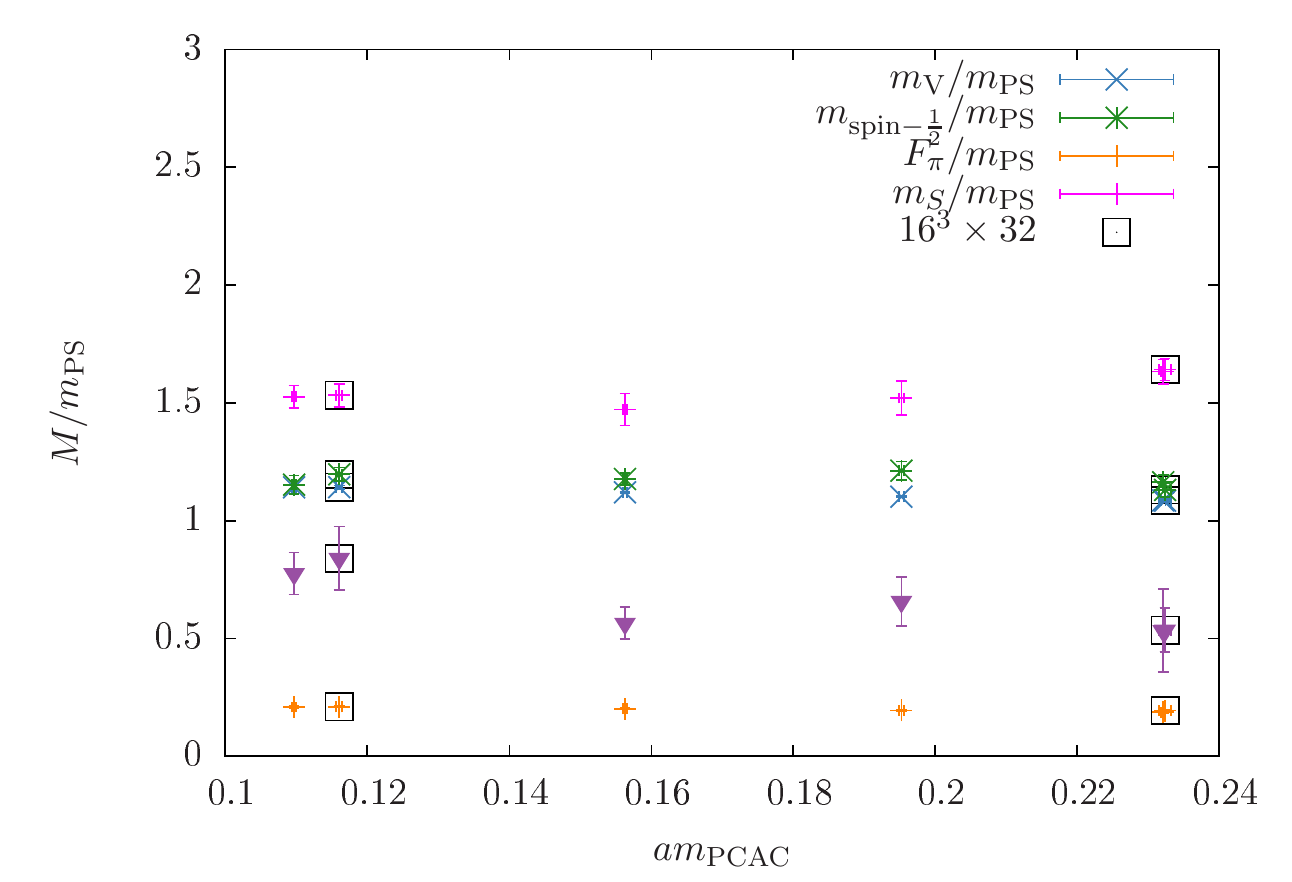}\label{ratiosnf32}}
 \caption{Mass spectrum of adjoint QCD with $N_f=3/2$ Dirac flavours at $\beta=1.5$ with the same lattice action as for the $N_f=2$ investigations, see Figure \protect\ref{mwtmass}.}
\end{figure}
In our most recent simulations we have started an investigation of the theory with three Majorana flavours, corresponding to $N_f=3/2$ Dirac flavours.
When integrating out an odd number of Majorana fermions, one obtains, in addition to the usual fermion determinant to the power of $N_f$, also the sign of the Pfaffian of the Dirac operator.
For our investigations of supersymmetric Yang-Mills theory we have developed methods to calculate the sign of the Pfaffian, but for our current parameters this contribution can be neglected.

We have investigated the mass spectrum of this theory and found that it is quite similar to the $N_f=2$ case. The scalar glueball is again the lightest state in the theory, and also the mass hierarchy is 
the same in both cases, see Figure \ref{massesnf32} and \ref{ratiosnf32}. The main exception that we have discovered so far is the spin-1/2 state: it changes from $\msh<\mps$ in the $N_f=2$ case to $\mps<\msh$ for $N_f=3/2$.

We have done a first preliminary investigation of the mass anomalous dimension from the mode number. For the two different coupling constants that we have currently considered, we obtain 
a value of $\gamma_\ast=0.40(5)$ at $\beta=1.5$, and $\gamma_\ast=0.32(5)$ at $\beta=1.7$. The values are hence larger than in the $N_f=2$ case, and apparently there are again some scaling corrections that would have to be 
included in order to obtain the final universal result.                                                          
\section{Conclusions: comparison of $N_f=1/2$ to $N_f=2$}
\begin{table}
\begin{center}
 
\begin{tabular}{|c|c|c|c||}
\hline
{\bf Theory} & {\bf scalar particle}  & {$\gamma_\ast$ small $\beta$}&{$\gamma_\ast$ larger $\beta$}\\
\hline
$N_f=1/2$ adj QCD (SYM) & part of multiplet & -- & --\\
\hline
$N_f=1$ adj QCD & light  & 0.92(1) &  $0.75(4)^\ast$\\
\hline
$N_f=3/2$ adj QCD & light  & $0.40(5)^\ast$ &  $0.32(5)^\ast$\\
\hline
$N_f=2$ adj QCD (MWT) & light  & 0.376(3) & 0.274(10) \\
\hline
\end{tabular}
\caption{The comparison of adjoint QCD with a different number of Dirac fermions $N_f$. The table states the relation of the scalar glueball to the rest of the bound state spectrum and shows the mass anomalous dimension.
The results indicated by a $\ast$ symbol are still in a preliminary state.\label{taball}}
\end{center}
\end{table}
Our results can be combined with other findings in order to obtain a more complete general picture for adjoint QCD with various numbers of fermions, see Table \ref{taball}.
The smallest possible number of fermions, $N_f=1/2$ corresponds to supersymmetric Yang-Mills theory. This theory has been considered in our earlier investigations \cite{Bergner:2015adz}.
It is a confining theory with supersymmetric multiplets of bound states. The lightest scalar is necessarily a component of the lightest multiplet and can hence not be 
separated from the rest of the spectrum.

One-flavour adjoint QCD has been considered in \cite{Athenodorou:2014eua}. Even with this rather small number of fermions, the theory seems to be in the conformal window. The lightest scalar particle is a 
scalar bound state and there is no clear signature of chiral symmetry breaking. 
Our results show how the mass anomalous dimension consistently decreases with the number of fermions. In addition also the gap between the lightest scalar bound state and the rest of the 
spectrum increases.
These bounds of the conformal behaviour can also be generalised to different $SU(N_c)$ and other representations, as detailed in \cite{Bergner:2015dya}. 

\section*{Acknowledgments}
This project is supported  by the John von Neumann Institute for Computing
(NIC) with grants of computing time. 
We also gratefully acknowledge the Gauss Centre for Supercomputing e.V.\ for funding this project by providing 
computing time on the GCS Supercomputer SuperMUC at Leibniz Supercomputing Centre.
Further computing time has been
provided by the computer cluster PALMA of the University of M\"unster.
\begin{spacing}{0.9}

\end{spacing}
\end{document}